\documentclass[aps,pra,reprint,groupedaddress]{revtex4-1}

\usepackage{graphicx}
\usepackage{amsmath}
\usepackage{bm}
\usepackage{braket}
\usepackage{hyperref}
\hypersetup{
    colorlinks=true,
    linkcolor=blue,
    filecolor=blue,      
    urlcolor=blue,
}

\begin{document}

\title{Structure and electron dynamics of planetary states of Sr below the Sr$^+$ $7d$ and $8p$ thresholds}

\author{M.\,G\'en\'evriez}
\affiliation{Laboratory of Physical Chemistry, ETH Z\"urich, CH-8093 Z\"urich}
\email[]{matthieu.genevriez@phys.chem.ethz.ch}
\author{C.\,Rosen}
\author{U.\,Eichmann}
\affiliation{Max-Born-Institute, 12489 Berlin, Germany}

\date{\today}

\begin{abstract}
In a combined experimental and theoretical study we investigate the $7dnl$ and $8pnl$ ($n\ge 11$, $l=9-12$) doubly-excited planetary states of Sr. The experimental spectrum was obtained using a five-photon resonant excitation scheme. The method of configuration interaction with exterior complex scaling was used to compute the energy-level structure and dynamics of the two highly excited electrons from first principles. Good quantitative agreement was obtained with the spectra we recorded, and the theoretical calculations shed light on their complex structure and the signatures of electron correlations therein. The two-electron probability densities we calculated reveal the strongly-correlated angular motion of the two electrons in the $7dnl$ and $8pnl$ planetary states, and confirm quantitatively the predictions of the frozen-planet approximation describing electron dynamics as the polarization of the fast inner electron by the electric field of the outer ``frozen'' electron.
\end{abstract}


\maketitle

\section{Introduction}

Electronic states of atoms and molecules in which two electrons are excited to
Rydberg states, the so-called double Rydberg states, have been extensively used
to study in great details many aspects of the three-body quantum mechanical
problem~\cite{aymar96,tanner00}. Because Rydberg electrons spend most of their time far away from the residual doubly-charged ion core, correlations between the two Rydberg electrons evolving in the Coulomb field of the core are magnified compared to residual interactions with the core electrons. Combined with the large density of accessible states, this makes electronic motion in double-Rydberg states rich and complex~\cite{richter93,camus94,tanner00} and may give rise to strongly-correlated dynamics in which each of the three Coulomb interactions plays an essential role, as is the case for the frozen-planet states theoretically predicted in the helium atom~\cite{richter91}.

Strontium atoms, along with the other alkaline-earth metals, are well-suited to study double Rydberg states experimentally because they possess only two valence electrons which can be photoexcited with conventional visible and ultraviolet lasers. Asymmetric double Rydberg states, in which the (approximate) principal quantum numbers associated with the two electrons are significantly different ($n_1 \ll n_2$), have been experimentally studied and characterized in Sr for a broad range of angular-momentum values~\cite{madden63,cooke78a,xu87,eichmann90,eichmann92,huang00,cohen01a,eichmann03,fields18}. Because they lie above the first ionization threshold, double-Rydberg states can decay rapidly by autoionization. Planetary configurations~\cite{percival77}, in which autoionization is suppressed and the two electrons orbit the nucleus in a solar-system-like manner, have been obtained~\cite{camus93,camus94,eichmann90,jones90,pisharody04}. In addition to their fundamental role in the development of multichannel theories~\cite{aymar87,aymar96}, core-excited Rydberg states, in which one of the two electrons is in a low-lying excited state of the ion core, recently gained particular interest in the context of quantum optics and quantum simulation~\cite{millen10,mukherjee11,teixeira20,madjarov20a,lehec21,wehrli19}. The possibility to manipulate the valence electron of the residual ion core within the Rydberg electron orbit offers a number of exciting perspectives to, \textit{e.g.}, detect~\cite{madjarov20a}, image~\cite{millen10} or trap~\cite{wilson19} Rydberg atoms.  

As the degree of excitation of both electrons increases ($n_1$ and $n_2$ large), spectra associated with double-Rydberg series becoming increasingly complex because the density of states increases and Rydberg-series interactions become ubiquitous. Several electrostatic models were developed to qualitatively describe the energy-level structure and two-electron wavefunctions of the high-angular-momentum double-Rydberg states measured in alkaline-earth-metal atoms~\cite{camus89,jones90,codling90,eichmann90,eichmann92,ostrovsky93a,huang94,heber97}. They describe the correlations between the two Rydberg electrons of such systems as the polarization of the fast inner electron by the electric field of the slow outer electron, the so-called frozen-planet approximation~\cite{eichmann90}. However, in contrast with the numerous studies carried out for the helium atom~\cite{tanner00,domke91,richter92,burgers95,czasch05,eiglsperger09,madronero08,gonzalez-melan20,grozdanov20a,wang21}, no quantitative theoretical information is available for alkaline-earth-metal atoms to date, in particular concerning the complex photoexcitation spectra recorded in the experiments. With one exception~\cite{wood94}, no attempt was made to calculate double-Rydberg states of Sr from first principles. With the development of the method of configuration interaction with exterior complex scaling (CI-ECS), accurate calculations that treat the interaction between the two electrons to all orders everywhere are now feasible~\cite{genevriez21}. It offers the possibility to (i) quantitatively analyze and assign experimental spectra, (ii) assess the validity of the aforementioned models, and (iii) investigate correlated electron dynamics in regions of phase space where these models are not applicable.

We present a joint experimental and theoretical study of planetary states of Sr located below the Sr$^+(7d)$ and Sr$^+(8p)$ limits. Experimental photoexcitation spectra were recorded from $5d_{5/2}16(l_2 \sim 10)$ states prepared by isolated-core excitation (ICE)~\cite{cooke78a}, as presented in Sec.~\ref{sec:experiment}. The large-scale CI-ECS approach we then used to calculate doubly-excited Rydberg states of Sr is described in Sec.~\ref{sec:theory}, along with the procedure employed to simulate experimental spectra. Experimental and theoretical spectra are presented and analyzed in Sec.~\ref{sec:results}. In the light of their good mutual agreement, reliable electronic densities derived from the CI-ECS calculations are then used to investigate electronic correlations and describe the two-electron collective motion in the planetary states under scrutiny. These results are further discussed in Sec.~\ref{sec:discussion} and analyzed in the light of existing models and theories describing double-Rydberg states.

\section{Experiment}\label{sec:experiment}

The experimental setup  and the resonant multiphoton laser excitation scheme used in the present experiment are identical  to the ones described in  \cite{huang00}. Strontium atoms in the ground state emanating from a resistively heated oven are excited in the presence of a constant electric
field to a Stark
 5$s16k$ ($m$=0)  by two excimer-laser-pumped dye lasers, which are parallel linearly polarized. $m$ is the magnetic quantum number and $k$ indicates a particular Stark state. After the excitation,
the electric field is switched off adiabatically within
1.5~$\mu$s and the Stark state is ideally converted into a single angular momentum eigenstate $5s16(l_{2}\sim10)$ \cite{freeman76}. As will be discussed in Sec.~\ref{sec:simulation}  non-adiabatic effects effectively result in an admixture of neighboring angular momentum states.
Applying the ICE technique, Sr atoms in the $5s16l_{2}$ state are further excited  via the $5p_{3/2}16l_2$ to the $5d_{5/2}16l_2$
state by another two dye lasers pumped by a second excimer laser. The two dye lasers are also parallel linearly polarized to ensure $\Delta m=0$ transitions only. Due to the high angular momentum involved, autoionization of the intermediate states is largely suppressed.   A fifth
dye laser is scanned in the energy range of the
$7d_{5/2}n'l'$ and $8p_{3/2}n''l''$ series. 
The doubly excited atoms either
autoionize, or are directly photoionized by the fifth dye laser. As detailed in section~\ref{sec:simulation} the resulting excited Sr$^+$ ions are further photoionized,  or ionized by a strong static field-pulse   to yield Sr$^{2+}$ ions, which are detected in our experiment. The Sr$^{2+}$ spectrum displayed in Fig.~\ref{fig:full_experimental_spectrum} is 
obtained by recording the Sr$^{2+}$ ion yield as a function of the
wavelength of the fifth laser. Calibration of the wavelength has been achieved by means of an optogalvanic Ar spectrum  and using the  known transition energies of Sr$^+$ ionic lines which appear in the spectra.   In
our time-of-flight spectrometer the static pulsed-field serves also  to sweep the Sr$^{2+}$ ions to the
multichannel plate ion detector.

\section{Theory}\label{sec:theory}

\subsection{Configuration interaction with exterior complex scaling}\label{sec:CI-ECS}

The CI-ECS method~\cite{genevriez21,genevriez19b} was used to calculate the
energies, widths and complex-scaled wave functions of the relevant Sr
doubly-excited states, and to determine the photoionization cross
sections and electronic densities presented in Sec.~\ref{sec:results}. We sketch below the parts of the method relevant to the present study. Atomic units are used throughout the rest of the section unless stated otherwise.

Within the CI-ECS approach, the two valence electrons of Sr are explicitly treated whereas the influence of the
electrons of the closed-shell Sr$^{2+}$ core is accounted for by an
$\ell$-dependent empirical model potential $V_{\ell}(r)$, whose parameters were
adjusted to reproduce the energies of the Sr$^+$
ion~\cite{aymar96}. Singly- and doubly-excited Rydberg states of
Sr are thus reduced described by the effective two-electron Hamiltonian,
\begin{align}
	\hat{H}(\bm{r_1},\bm{r_2}) =& -\frac12\bm{\nabla}_1^2 - \frac12\bm{\nabla}_2^2 + V_{l_1}(r_1) + V_{l_2}(r_2) + \frac{1}{r_{12}} \nonumber\\
	&+ V^\text{SO}_{s_1l_1j_1}(r_1) + V^\text{SO}_{s_2l_2j_2}(r_2) + V^{(2)}_{\text{pol}}(\bm{r_1}, \bm{r_2}) .
	\label{eq:two_electron_hamiltonian}
\end{align}
The two electrons have positions with respect to the nucleus given by
$\bm{r_1}$ and $\bm{r_2}$, and are described by the independent-electron orbital-, spin-
and total-angular-momentum quantum numbers $\ell, s$ and $j$. The distance between the two electrons is given by $r_{12} =
|\bm{r}_1 - \bm{r}_2|$.
$V^\text{SO}_{s \ell j}(r)$ is the spin-orbit interaction between a valence
electrons and the screened nucleus and $V^{(2)}_{\text{pol}}(\bm{r_1},
\bm{r_2})$ represents bielectronic core polarization. Their expressions are
given, \textit{e.g.}, in Ref.~\citenum{genevriez21}.

The time-independent Schr\"odinger equation associated with the Hamiltonian in
Eq.~\eqref{eq:two_electron_hamiltonian} is solved, in a
configuration-interaction (CI) manner, using a large basis of two-electron functions
built from antisymmetrized products of two one-electron spin-orbitals of the
Sr$^+$ ion. Continuum states and resonances are accounted for with exterior
complex scaling~\cite{simon79}, following which the radial coordinate of each
electron is rotated into the complex plane after a certain ECS radius $r_0$,
\begin{equation}
	r \to \begin{cases}
		r &\qquad \text{if} \quad r < r_0 ,\\
		r_0 + (r - r_0)\text{e}^{\text{i}\theta} &\qquad \text{if} \quad r \ge r_0 .
	\end{cases}
	\label{eq:ECS_contour}
\end{equation}
As a result of ECS, the wavefunctions associated with resonances, which are not square integrable for real $r$ values, become exponentially damped at large distances and the entire spectrum of bound and resonance radial wave functions can be represented by a discrete set of square-integrable functions.

In practice, the numerical one-electron radial functions used to build the two-electron CI wavefunctions are calculated along the ECS
contour~\eqref{eq:ECS_contour} using a finite-element
discrete-variable-representation (FEM-DVR) method~\cite{rescigno00}. The
calculation parameters are listed in Table~\ref{tab:FEM-DVR_parameters}, and
were chosen as described in Ref.~\citenum{genevriez21}. Because, in a
doubly-excited state, only one of the two electrons of can autoionize,
the description of the second electron can be limited to a relatively compact set of bound one-electron orbitals of Sr$^+$. This
dramatically reduces the computational cost of solving the two-electron
Schr\"odinger equation and makes it possible to treat accurately high-lying doubly-excited states. In the present calculation, approximately 80\,000 two-electron
wave functions were built from (i) all spin-orbitals with energies lower than
the Sr$^+(11s)$ state and (ii) the complete set of spin-orbitals associated with
the 208 radial functions generated by the FEM-DVR calculation. The ECS radius $r_0 = 150\ a_0$ is chosen such that the amplitudes of the radial functions of all states below the Sr$^+(11s)$ state are negligible at $r_0$ and beyond.

\begin{table}
	\begin{ruledtabular}
	\begin{tabular}{c c c c}
	Element & $i=1$ & $i=2$ & $i=3$\\
	$[r_{i}, r_{i+1}]$ & $[0, 10]\ a_0$ & $[10, 150]\ a_0$ & $[150, 1350]\ a_0$ \\
	$N_i$ & 70 & 60 & 80 \\
	$\theta_i$ & 0 & 0 & $10^\circ$ \\		
	\end{tabular}
	\end{ruledtabular}
	\caption{FEM-DVR parameters used in the present calculations. The $i$-th element spans radial distances from $r_i$ to $r_{i+1}$ and contains $N_i$ grid points. The complex-scaling angle in this element is $\theta_i$.}
	\label{tab:FEM-DVR_parameters}
\end{table}

The complex-scaled two-electron Hamiltonian matrix
is constructed using the large CI basis set and iteratively diagonalized in the
relevant energy region with a Lanczos algorithm adapted to complex-symmetric matrices. Resonances associated with doubly-excited states have complex eigenvectors and eigenvalues $E - \mathrm{i}\Gamma/2$ giving the energies $E$ and autoionization widths $\Gamma$ of these states. We ensured that all relevant eigenvalues are converged to better than 0.5~$\mu$Hartree (0.1 cm$^{-1}$) with respect to the number of grid points, the number of basis functions and the complex-scaling angle.

The present calculations take into account the spin-orbit interaction for \emph{both} electrons and are carried out using the $jj$ coupling scheme. For the planetary states considered in the present work, the spin-orbit interaction for the outer electron is very small and these states are commonly described using the $jK$ coupling scheme. The transformation between the $jj$ and $jK$ coupling schemes is given by~\cite{cowan81}
\begin{align}
&\braket{l_1 s_1 j_1 l_2 K (s_2 J) | l_1 s_1 j_1 l_2 s_2 j_2 J} = \nonumber\\
&(-1)^{j_1+l_2+s_2+J}\sqrt{(2K+1)(2j_2+1)}
\begin{Bmatrix}
j_1 & l_2 & K \\
s_2 & J & j_2
\end{Bmatrix} .	
\label{eq:jK_jj_transformation}
\end{align}

Total photoionization cross sections are calculated from the CI-ECS wavefunctions following the procedure of Rescigno and McKoy~\cite{rescigno75} (see also Ref.~\citenum{genevriez19b}). The initial Sr($5dn_2l_2$) states are already autoionizing resonances lying above the first ionization threshold, but are treated as nondecaying states because of their low autoionization rates ($< 2 \times 10^{-9}$~a.u., \textit{i.e.}, lifetimes $>12$~ns). Partial photoionization cross sections were calculated from the complex-scaled wavefunction in the interior region ($r \le r_0$) using an approach proposed by Carette \textit{et al.}~\cite{carette13} and Miheli{\v c}~\cite{mihelic18}. We start from the driven Schr\"odinger equation,
\begin{equation}
		(E_0 + \omega - \hat{H}(\bm{r_1},\bm{r_2})) \ket{\Psi_1} = \hat{D} \ket{\Psi_0} ,
		\label{eq:driven_SE}
\end{equation}
where $\ket{\Psi_0}$ is the wave function of the initial state with energy
$E_0$, $\omega$ is the photon angular frequency and $\hat{D}$ is the transition
dipole operator. In the length gauge, used in the present work,
$\hat{D}=-\bm{\hat{\epsilon}}\cdot (\bm{r_1} + \bm{r_2})$ where
$\hat{\epsilon}$ is the polarization vector. The solution to
Eq.~\eqref{eq:driven_SE} within the ECS framework is given by
\begin{equation}
	\ket{\Psi_1^\theta(\omega)} = \sum_i \ket{i^\theta}\frac{\braket{i^\theta | \hat{D} | \Psi_0}}{E_0 - E_i^\theta + \omega} ,
\end{equation}
where the summation runs over all ECS states $\ket{i^\theta}$ obtained by
diagonalization of the complex-scaled two-electron Hamiltonian. Within the
interior region ($r < r_0$), $r$ is real and the wave function
$\ket{\Psi_1^\theta(\omega)}$ is identical to the one without complex scaling.
In the exterior region ($r \ge r_0$), $r$ possesses an imaginary component
such that the physical significance of $\ket{\Psi_1^\theta(\omega)}$ and its use to extract physically-relevant information is significantly complicated. Therefore, partial cross sections are best obtained from $\ket{\Psi_1^\theta(\omega)}$ in the interior region just before $r_0$, where we assume that it has already reached its asymptotic, Coulomb-type behavior. The function $\ket{\Psi_1^\theta(\omega)}$ is projected onto channel functions, which describe all but the radial motion of the photoelectron~\cite{aymar96}, yielding the radial function $P_{\alpha}(r_2 ; \omega)$. $\alpha$ stands for the quantum numbers $n_1, \ell_1, j_1, \ell_2, j_2, J$ and $M$ which define a channel. Provided that $r_0$ is sufficiently large, $P_{\alpha}(r_2 ; \omega)$ can be represented for $r_2 \lesssim
r_0$ by a linear combination
of regular $F_{\ell_2}^E$ and irregular $G_{\ell_2}^E$ Coulomb functions,
\begin{equation}
	P_{\alpha}(r_2 ; \omega) \sim A_\alpha(\omega) \left[ F_{\ell_2}^E(-1/k, kr_2) + \mathrm{i}G_{\ell_2}^E(-1/k, kr_2) \right] .
	\label{eq:asymptotic_form}
\end{equation}
The imaginary part of the right hand side of Eq.~\eqref{eq:asymptotic_form} is fitted to $\text{Im}\left(P_\alpha\right)$, calculated from the CI-ECS wavefunction, in a linear least-squares fit with the amplitude $A(\omega)$ as a fit parameter. The partial cross section $\sigma_\alpha(\omega)$ at a photon angular frequency $\omega$ is then obtained from $A(\omega)$ using
\begin{equation}
	\sigma_\alpha(\omega) = \frac{4 \omega}{c} \left| A_\alpha(\omega)\right|^2 .
\end{equation}

Electronic correlations are investigated in their finest details by computing
and analyzing the two-electron density (see, \textit{e.g.}, Refs.~\cite{ezra83,richter92}). Because of ECS, the radial coordinates are rotated into the complex plane and the density cannot be straightforwardly computed from complex-scaled two-electron wavefunction everywhere. Deducing the real-$r$ wavefunction from its complex-rotated counterpart, a process known as backscaling, is possible for standard complex scaling ($r_0 = 0$) but suffers from severe numerical instabilities~\cite{buchleitner94}. It is also unclear whether it can be extended to ECS. Instead, just as for computing partial photoionization cross sections, we use the fact that for $r \le r_0$ the radial coordinates are real and the functions $|i^\theta\rangle$ can be identified to their non-complex-scaled versions. Provided that we restrict ourselves to $r_1, r_2 \le r_0$, the electronic density $\rho(\bm{r}_1, \bm{r}_2)$ can thus be calculated as usual. Integration of $\rho$ over the Euler angles describing the orientation the electron-electron-nucleus triangle reduces the 6-dimensional coordinate space to the three internal variables $r_1$, $r_2$ and
$\theta_{12}$ relevant for electron-electron correlations. The corresponding reduced two-electron density is calculated as

\begin{align}
	\rho_i(r_1&, r_2, \cos\theta_{12}) =\\
	&\braket{i^\theta | \delta(r_1 - r'_1)\delta(r_2 - r'_2)\delta(\cos\theta_{12} - \cos \theta'_{12}) | i^\theta} , \nonumber
\end{align}
where $\delta(x)$ are Dirac delta functions. The expansion of $\delta(\cos\theta_{12} - \cos \theta'_{12})$ in terms of Legendre polynomials is used for the calculations~\cite{warner80}.

\subsection{Simulation of experimental spectra}\label{sec:simulation}

\begin{figure}
	\includegraphics[width=\columnwidth]{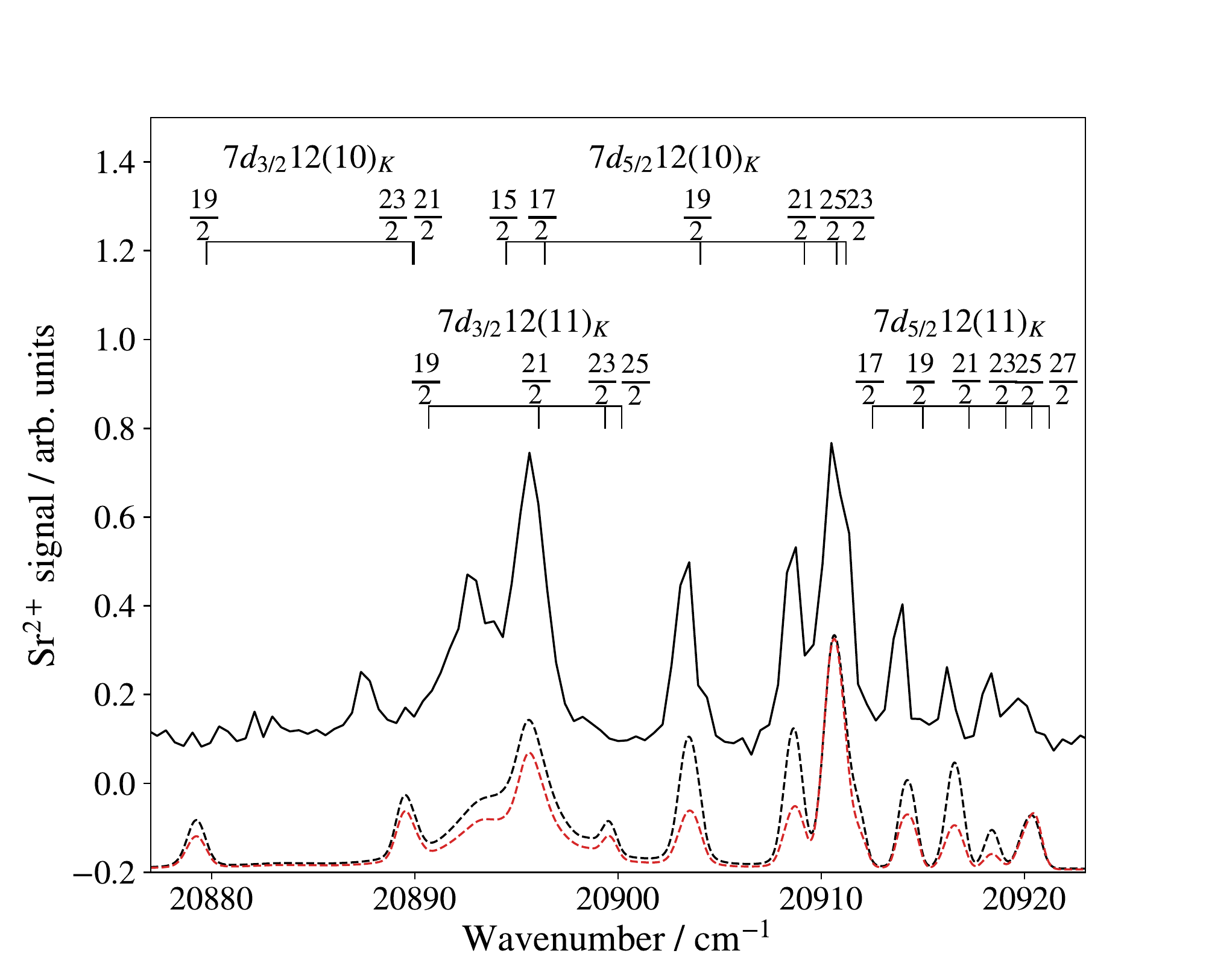}
	\caption{Experimental (full line) and theoretical (dotted lines) Sr$^{2+}$
	spectra in the region of $7d_{3/2, 5/2}12(l'_2)$ resonances. The red dotted
	line corresponds to a spectrum calculated from the partial photoionization cross
	section to states energetically higher than Sr$^+(6d)$ only, whereas the black dotted line was
	obtained from the total cross section. The theoretical spectra were offset by -0.2 along the vertical axis and
	scaled by a factor of 2 for clarity.}
	\label{fig:7d_n12_zoom}
\end{figure}

To reproduce the experimental spectra, the production dynamics of Sr$^{2+}$ ions following excitation to doubly-excited states must be elucidated. First, $(7d_{j'_1}n'_2(l'_2)_{j'_2})_{J'}$ or $(8p_{j'_1}n'_2(l'_2)_{j'_2})_{J'}$ states are photoexcited from $(5d_{5/2}16(l_2)_{j_2})_{J}$ states. Sr$^{2+}$ ions are then detected following,
predominantly, (i) autoionization and subsequent photoionization of the
residual Sr$^+$ ion or (ii) photoionization of the core electron and subsequent
field ionization of the Sr$^+$ ion in a Rydberg state. In process (i), the
final state of the ion after autoionization is important. If it is
energetically too low ($\le \text{Sr}^+(6d_{5/2})$), the photon energy is not
sufficient to photoionize the ion and Sr$^{2+}$ ions are not detected. The
photoionization cross section is also different for each
Sr$^+$ final state. To assess the importance of these effects, we
calculated partial photoionization cross sections to each Sr$^{+}$ final state and weighted them by the corresponding Sr$^+$ photoionization cross section. The corresponding spectra are very
similar to the spectra obtained including all Sr$^+$ final states and without
weighting (see red and black dotted lines in Fig.~\ref{fig:7d_n12_zoom}). Process (ii) was studied in detail by
Rosen \textit{et al.}~\cite{rosen99}. Within the independent-electron
approximation, the photoionization of the core electron is independent of the
state of the Rydberg electron, and the field ionization of the Sr$^+$ Rydberg
ion is also independent of the particular Rydberg state for sufficiently large
field strengths. The competition between processes (i) and (ii) depends in a
complicated manner on the photoexcitation and autoionization dynamics. For the
sake of simplicity, the theoretical spectra presented below are thus obtained
from the total photoionization cross section of $(5d_{5/2}16(l_2)_{j_2})_{J}$ states unless stated otherwise.

The calculated photoionization cross sections are convolved by a Gaussian function with a full-width-at-half-maximum of 0.25~cm$^{-1}$ to account for the finite laser bandwidth in the experiment. Theoretical spectra are obtained from the convolved cross section $\sigma_\text{c}(\omega)$ using
\begin{equation}
	S(\omega) = \sum_i P_i \left[\ln\left( p_\text{sat} \sigma^i_{\text{c}}(\omega)\right) + \gamma - \text{Ei}\left( -p_\text{sat} \sigma^i_{\text{c}}(\omega)\right)\right] ,
\end{equation}
where the coefficients $P_i$ are the populations of the various
$(5d_{5/2}16(l_2)_{j_2})_{J}$ initial states. $\gamma$ is the Euler constant,
$\text{Ei}(x)$ is the exponential integral function and $p_\text{sat}$ is a
parameter adjusted to visually reproduce experimental spectra. This formula
accounts exactly for both saturation and interaction-volume effects in the case
of a Gaussian laser beam and a large atomic beam~\cite{antoine96}.

In the experiment, Sr atoms are prepared in
$\left(5d_{5/2}16(l_2)_{j_2}\right)_J$ states by ICE from bound singlet $5s16(l_2)$ Rydberg states, for which the $LS$ coupling scheme is appropriate. In order to reproduce experimental spectra, the relative populations $P_i$ of the various $\left(5d_{5/2}16(l_2)_{j_2}\right)_J$ states must be estimated. A complete description of the preparation process using CI-ECS is computationally very demanding, since many eigenvalues and eigenvectors must be obtained from the large Hamiltonian matrix. Instead, we obtained an initial estimate using the ICE approximation~\cite{bhatti81} which relies on the independent-electron approximation. We treated the $5p_{3/2}16l_2$ and $5d_{5/2}16l_2$ weakly autoionizing  doubly-excited states as nondecaying states, \textit{i.e.}, we neglected the admixture of any continuum state, because as mentioned in Sec.~\ref{sec:CI-ECS} their autoionization rates are small on the timescale of the experiment.

The initial $5s16(l_2)$ singlet states are projected onto the $jj$-coupled
$(5s_{1/2}16(l_2)_{j_2})_{J''=l_2}$ states using standard angular
momentum algebra~\cite{cowan81}. The photoexcitation cross section from an
initial $|n_1 l_1 j_1 n_2 l_2 j_2 J M\rangle$ state can be written, in the ICE
approximation, as
\begin{widetext}
	\begin{align}
	\sigma_\text{ICE}(\omega) = \frac{4\pi^2\omega}{c} &\left[J, J', j_1, j'_1\right]
	\begin{pmatrix}
	J' & 1 & J \\
	-M' & q & M
	\end{pmatrix}^2
	\begin{Bmatrix}
	j_2 & j'_1 & J' \\
	1   & J  & j_1
	\end{Bmatrix}^2
	\begin{Bmatrix}
	l'_1  & 1/2 & j'_1 \\
	j_1 & 1  & l_1
	\end{Bmatrix}^2
	\label{eq:ice_preparation}\\
	&
	\times|\langle n'_1 l'_1 j'_1 | \hat{\epsilon}\cdot \vec{r}_1 | n_1 l_1 j_1 \rangle|^2 |\langle n'_2 l_2 j_2 | n_2 l_2 j_2 \rangle|^2 A_{n'_1l'_1j'_1}^{l_2j_2 J'}(\omega)
	\nonumber ,
\end{align}
\end{widetext}
where $\omega$ is the photon angular frequency and $c$ is the speed of light. We use the standard notation $[j] = 2j+1$. The first term on the second line is the square of the transition dipole moment
for the excitation of the bare Sr$^+$ ion from the $n_1(l_1)_{j_1}$ state to
the $n'_1(l'_1)_{j'_1}$ state. The second term describes the overlap between
the initial and final Rydberg-state wavefunctions, and is well approximated by
the $\text{sinc}^2$-type function~\cite{bhatti81}
\begin{equation}
	|\langle n'_2 l_2 j_2 | n_2 l_2 j_2 \rangle|^2 = \frac{4(n'_2n_2)^4}{n_2^3(n_2+n'_2)^2}\text{sinc}^2\left(n'_2 - n_2\right) .
	\label{eq:overlap_integral}
\end{equation}
The last term in Eq.~\eqref{eq:ice_preparation} is the spectral density of the $n'_1(l'_1)_{j'_1}(l_2)_{j_2}J'$ doubly-excited Rydberg states. In the absence of electron correlations, it is given by a
series of narrow Lorentzian functions located at the energies of the Rydberg states.
Because $l_2$ is large, the interaction between the two electrons during the
preparation process is very small and, in particular, the quantum defects are assumed to be
negligible ($n'_2 = n_2$ and $E_{n'_2} = E_{n_2}$). Consequently, the last two
terms are nonzero only when $\omega$ corresponds to the $n_1l_1j_1 -
n'_1l'_1j'_1$ transition frequency of the bare Sr$^+$ ion,  and are independent
of the values of $n_2$ and $l_2$. In other words, we assume that the Rydberg electron is a
spectator of core excitation and does not undergo shake-up or shake-down.
Under these assumptions, the relative excitation efficiencies from, first, the
$(5s_{1/2}16(l_2)_{j_2})_{J''=l_2}$ states to the $(5p_{3/2}16(l_2)_{j_2})_{J'}$ states and, second, from the $(5p_{3/2}16(l_2)_{j_2})_{J'}$ states to the
$(5d_{5/2}16(l_2)_{j_2})_{J}$ states, can be directly computed using
Eq.~\eqref{eq:ice_preparation}. Because lasers with parallel linear polarizations were used in the experiment, $M=q=0$. In a second step, the calculated values were
manually adjusted by visually comparing theoretical and experimental spectra.

\begin{table}
\begin{ruledtabular}
	\begin{tabular}{c c c c}
		& $J=l_2 - 2$ & $J=l_2$ & $J=l_2 + 2$ \\
		$(5d_{5/2}16(l_2)_{l_2 - 1/2})_J$ & 0.17 & 0.19 & 0.09 \\
		$(5d_{5/2}16(l_2)_{l_2 + 1/2})_J$ & 0.08 & 0.20 & 0.27 \\[0.15cm]
		\hline
		$l_2=9$ & $l_2=10$ & $l_2=11$ & $l_2=12$ \\
		0.44 & 0.4 & 0.08 & 0.08 \\
	\end{tabular}
\end{ruledtabular}
\caption{Populations of the various $(5d_{5/2}16(l_2)_{j_2})_{J}$ initial states.}
\label{tab:populations}
\end{table}

Because of nonadiabatic effects during Stark switching, the atoms are not
necessarily prepared in a $6s16(l_2)$ state with a single value of $l_2$. We carried out calculations for a number of $l_2$ values and added them to obtain the total spectrum, with weights corresponding to the respective $l_2$ populations. The $l_2$ weights and populations $P_i$ are summarized in Table~\ref{tab:populations}.

\begin{figure}
	\includegraphics[width=\columnwidth]{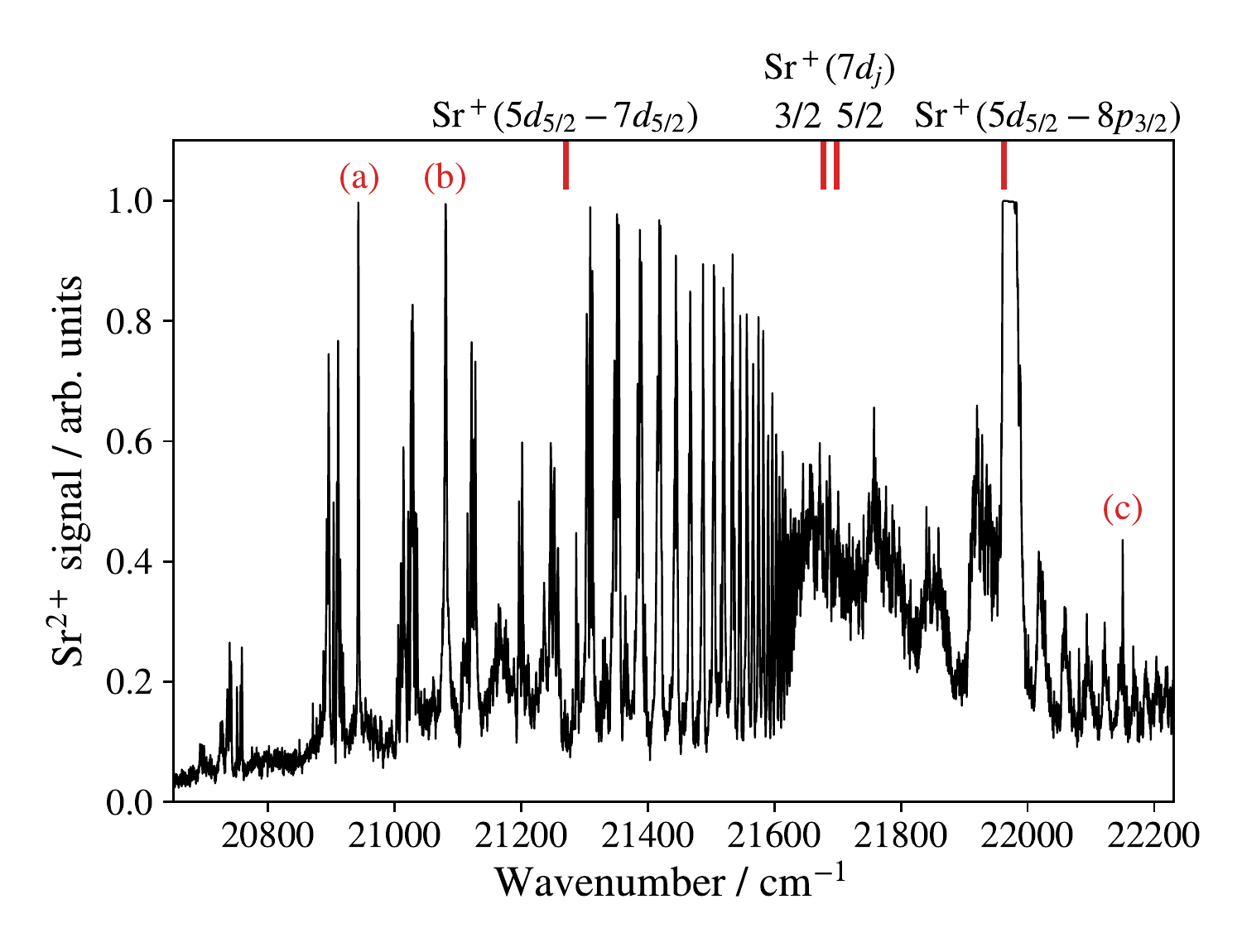}
	\caption{Experimental Sr$^{2+}$ spectrum recorded from $(5d_{5/2}16(l_2)_{j_2})_{J}$ states. The vertical lines on the top horizontal axis indicate the wavenumbers of the Sr$^+(5d_{5/2} - 7d_{5/2})$ non-dipole allowed isolated-core resonance, of the Sr$^{+}(7d_{j=3/2, 5/2})$ ionization thresholds, and of the Sr$^+(5d_{5/2} - 8p_{3/2})$ isolated-core resonance, respectively. The labels (a), (b) and (c) indicate the wavenumbers of the one-photon ionic transitions Sr$^+(6s_{1/2} - 7p_{1/2})$ and Sr$^+(6s_{1/2} - 7p_{3/2})$, and of the two-photon ionic transition Sr$^+(5p_{3/2}-7p_{3/2})$.}
	\label{fig:full_experimental_spectrum}
\end{figure}

\begin{figure*}
	\includegraphics[width=\textwidth]{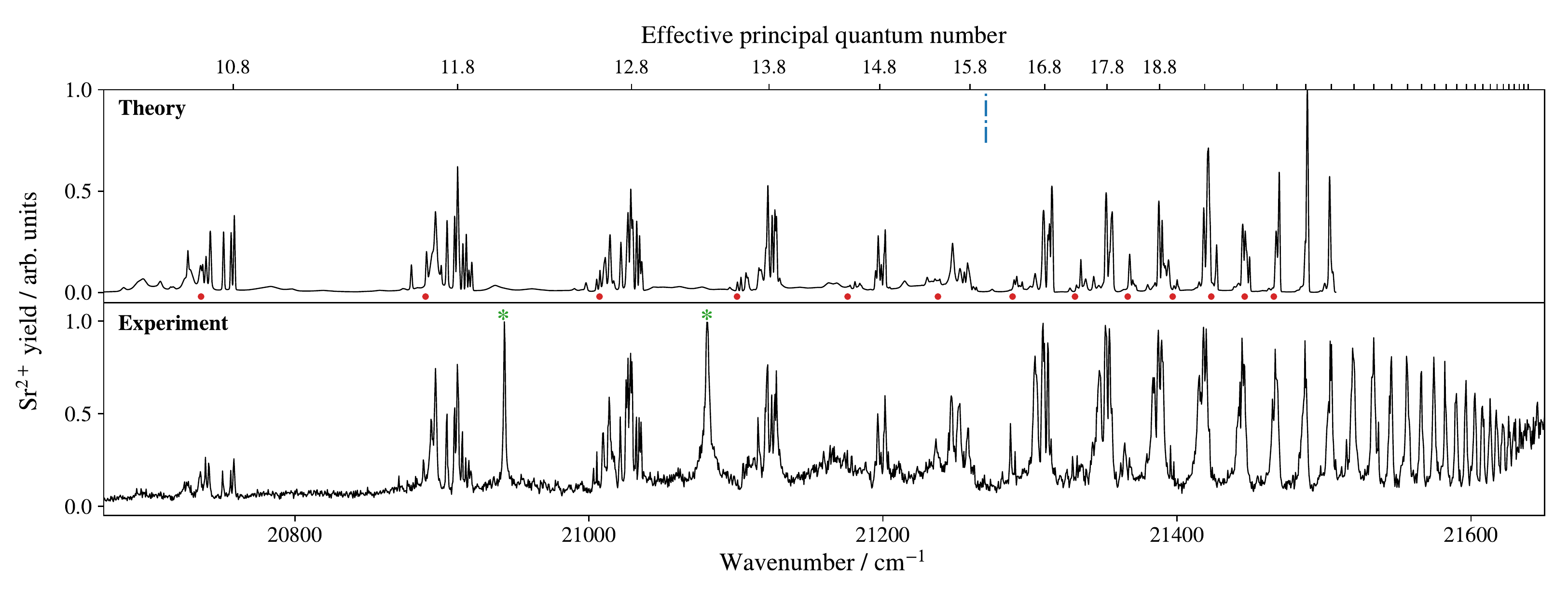}
	\caption{Theoretical (top) and experimental (bottom) Sr$^{2+}$ spectra from $(5d_{5/2}16(l_2)_{j_2})_{J}$ states ($l_2 = 9-12$) in the vicinity of the Sr$^+(5d_{5/2} - 7d_{5/2})$ transition (vertical chained line). The asterisks in the lower panel denote the Sr$^+(6s_{1/2}-7p_{1/2, 3/2})$ resonances. Effective principal quantum numbers relative to the Sr$^{+}(7d_{5/2})$ threshold  are shown on the upper horizontal axis. The circles in the upper panel show the position of states with effective principal quantum numbers starting from 10.8 and converging to the Sr$^{+}(7d_{3/2})$ threshold.}
	\label{fig:7d_region}
\end{figure*}

\section{Results}\label{sec:results}

\subsection{Photoionization spectra}

The experimental Sr$^{2+}$ overview spectrum we recorded is shown in Fig.~\ref{fig:full_experimental_spectrum}. The regions below and above the Sr$^{+}(7d)$ ionization threshold are discussed separately in the following. The theoretical and experimental Sr$^{2+}$ spectra in the first region, where $7dnl$ doubly-excited states can be excited, are compared in Fig.~\ref{fig:7d_region}. The theoretical spectrum is truncated to Rydberg
states with $n \le 24$, corresponding to the highest $n$ value accurately
described by our basis set. The spectrum was displaced along the horizontal axis by the difference between the calculated and experimental~\cite{kramida20} wavenumbers of the Sr$^+(5d_{5/2} - 7d_{5/2})$ transition ($26.42$~cm$^{-1}$). In the experiment, the third and fourth lasers used to prepare $5d_{5/2}nl$ states are tuned to the positions of the ionic resonances, thus assuming zero quantum defects of the outer Rydberg electron (see Sec.~\ref{sec:simulation}), whereas in the theoretical calculations the initial $(5d_{5/2}16(l_2)_{j_2})_{J}$ states possess a small positive quantum defect and thus lie slightly lower in energy ($\sim -0.8$~cm$^{-1}$). This results in an offset of $\sim -0.8$~cm$^{-1}$ between the experimental and theoretical wavenumbers for the fifth laser, which we correct by shifting the theoretical spectrum along the horizontal axis accordingly.
Overall, the agreement between theory and
experiment is very good considering the complexity of the experiment.
Discrepancies can be attributed to uncertainties in the modeling of the
Sr$^{2+}$ production (see Sec.~\ref{sec:simulation}), variations of the laser pulse energy or of the detection efficiency in the experiment, inaccuracies of the
calculation and, in particular, of the model potential, missing initial states
or inaccurate initial populations. In many cases a quantitative level of
agreement is reached, which was never obtained for such high-lying
doubly-excited states before~\cite{wood94}.

The spectra show a regular Rydberg-series progression, with an associated
quantum defect of $\sim 0.2$, converging to the Sr$^+(7d_{5/2})$ threshold. A
weaker series with approximately the same quantum defect and converging to the
other spin-orbit component ($7d_{3/2}$) is marked by the circles. For
singly-excited Rydberg states and core-excited Rydberg states, the quantum defects of high-$l$ Rydberg series ($l=9-12$ in
the present case) are very small. The interaction between the core- and Rydberg
electrons is indeed strongly reduced by the large centrifugal barrier which
prevents the penetration of the Rydberg electron in the core region.
Conversely, the large ($\sim 0.2$) quantum defects observed in the present
spectra indicate much stronger correlations between the two electrons, which
will be analyzed in more details in Sec.~\ref{sec:electronic_structure}.
Correlations are in fact fully responsible for the spectra shown in
Fig.~\ref{fig:7d_region}. In an ICE scheme, where electron correlations are
neglected, the excitation of $(7d_{j'_1}n'_2(l'_2)_{j'_2})_{J'}$ states from
$(5d_{5/2}16(l_2)_{j_2})_J$ states is forbidden because dipole selection
rules prevent the $5d \to 7d$ core excitation. Huang \textit{et al.} observed a
similar violation of ICE predictions for higher $l$ states ($l=13,
15$)~\cite{huang00}, which was also observed in CI-ECS
calculations~\cite{genevriez21}. In the present case, inspection of the CI
coefficients reveals that, for example, states with predominant
$(7d_{5/2}n'_2(10)_{j'_2})_{J'}$ character are mixed at the level of 1\% with
$(8p_{3/2}n''_2(9, 11)_{j_2})_{J'}$ and $(6f_{7/2}n''_2(9, 11)_{j_2})_{J'}$ states. ICE to the two latter states is allowed by dipole selection rules and, because the transition dipole moment from the $5d_{5/2}$ state of Sr$^+$ to the $6f_{7/2}$ state is much larger, it is the $(6f_{7/2}n''_2(9, 11)_{j_2})_{J'}$ character that provides the major intensity contribution to the photoexcitation spectrum.

\begin{figure*}
	\includegraphics[width=\textwidth]{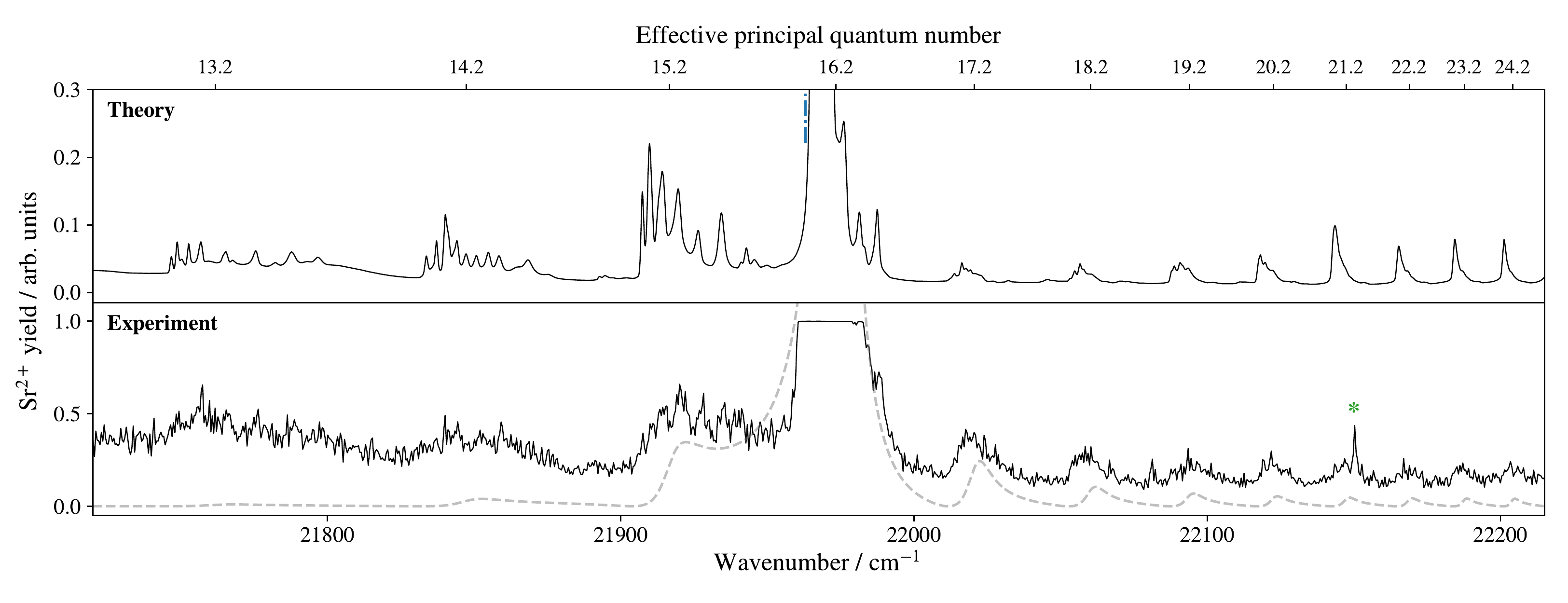}
	\caption{Theoretical (top) and experimental (bottom) Sr$^{2+}$ spectra from
	$(5d_{5/2}16(l_2)_{j_2})_{J}$ states ($l_2 = 9-12$) in the vicinity of
	the Sr$^+(5d_{5/2} - 8p_{3/2})$ transition. The vertical chained line in the upper panel shows the position of the Sr$^+(5d_{5/2} - 8p_{3/2})$ transition. The asterisk in the lower panel denotes the Sr$^+(5p_{3/2}-7p_{3/2})$ two-photon resonance. Effective principal quantum numbers relative to the Sr$^{+}(8p_{3/2})$ threshold  are shown on the upper horizontal axis. The dashed gray line in the lower panel shows a typical prediction of the ICE model.}
	\label{fig:8p_region}
\end{figure*}

Line intensities in the spectra show a significant decrease around
21250~cm$^{-1}$, a region where the principal quantum number of the outer
electron is approximately conserved upon excitation ($n'_2 \sim 16$, see upper
bar in Fig.~\ref{fig:7d_region}). This fact was already observed and explained
by Huang \textit{et al.}~\cite{huang00}. If we consider an initial state
described by a single configuration $(5d_{5/2}16(l_2)_{j_2})_J$, the largest
contributions to the transition dipole moments to $(7d_{5/2}16(l'_2)_{j'_2})_{J'}$ states come from the mixing of the latter states with configurations in which
the Rydberg electron is in the same configuration as the initial state but the
core-electron configuration has changed, \textit{e.g.},
$(6f_{7/2}16(l_2)_{j_2})_{J'}$ or $(8p_{3/2}16(l_2)_{j_2})_{J'}$. If one assumes
that the Rydberg electron is well described by a hydrogenic wavefunction, a
rather good approximation for high-$l_2$ states, then electrostatic interactions coupling the two configurations vanish for $n_2 = n'_2 =
16$~\cite{pasternack62}. This means that, in first approximation, the
$(7d_{5/2}16(l'_2)_{j'_2})_{J'}$ states possess only little
$(6f_{7/2}16(l_2)_{j_2})_{J'}$ or $(8p_{3/2}16(l_2)_{j_2})_{J'}$ character and
thus cannot be efficiently excited. This is no longer the case for noninteger
values of $n_2$ and $n'_2$ or $n_2 \neq n'_2$, which explains why the line
intensity does not completely vanish at $n'_2 \sim 16$ and why lines away from the $n'_2 \sim 16$ region are more intense. 

Each $n'_2$ band in the spectra possesses a complex substructure. A detailed
view of the $n'_2 \sim 12$ region is shown in
Fig.~\ref{fig:7d_n12_zoom}. The agreement between theory (black dotted line) and
experiment (full line) is very good and permits the assignment of each line to
one or a few states with predominant $(7d_{j'_1}12(10-12)_{j'_2})_{J'}$ character. The spin-orbit interaction for the Rydberg
electron is very small and levels with different $j'_2$ values but identical
quantum numbers otherwise are degenerate on the scale of the figure. Therefore,
states were labeled in $jK$ coupling using Eq.~\eqref{eq:jK_jj_transformation},
as shown by the assignment bars. States with $l'_2=9$ are broad and do not yield clear spectral features in the spectrum.

The theoretical and experimental Sr$^{2+}$ spectra recorded from
$(5d_{5/2}16(l_2)_{j_2})_J$ states and for photon energies above the Sr$^{+}(7d)$ threshold, a region where $8pnl$
doubly-excited Rydberg states are excited, are compared in
Fig.~\ref{fig:8p_region}. As for the $7dnl$ region, the theoretical spectrum was displaced along the horizontal axis by the difference between the calculated and experimental wavenumbers of the Sr$^+(5d_{5/2} - 8p_{3/2})$ transition ($22.0$~cm$^{-1}$, see paragraph below) and by an additional $-0.8$~cm$^{-1}$. Agreement between theory and experiment is satisfactory. The line positions are well reproduced by theory, however lines
tend to be broader in the experimental spectrum. This may be caused by
inaccuracies of the model potential, by the truncation of the basis set, or
by the fact that initial states with lower $l_2$ values are populated by
nonadiabatic effects in the experiment. This results in $8pnl'_2$ states with
lower $l'_2$ values being excited, which possess larger autoionization rates and
give rises to broader spectral lines. The peak corresponding to an effective
principal quantum number of $n'_2 \sim 16.2$ is heavily saturated under the
present experimental conditions.

Rydberg series with a negative quantum defect of $\sim -0.2$ are observed in Fig.~\ref{fig:8p_region} and
converge to the Sr$^+(8p_{3/2})$ threshold. The position of this threshold
obtained by extrapolation of the Rydberg formula to $n \to \infty$ is
$75335.7(7)$~cm$^{-1}$ relative to Sr$^+(5s)$, a value $23.9$~cm$^{-1}$ higher than the reference data from NIST~\cite{kramida20}. Discrepancies of similar magnitudes were observed earlier by Lange \textit{et al.}~for Sr$^+(7p_{1/2, 3/2})$~\cite{lange91}. We therefore used, in the present article, the value determined from our spectra.

\begin{figure}
	\includegraphics[width=\columnwidth]{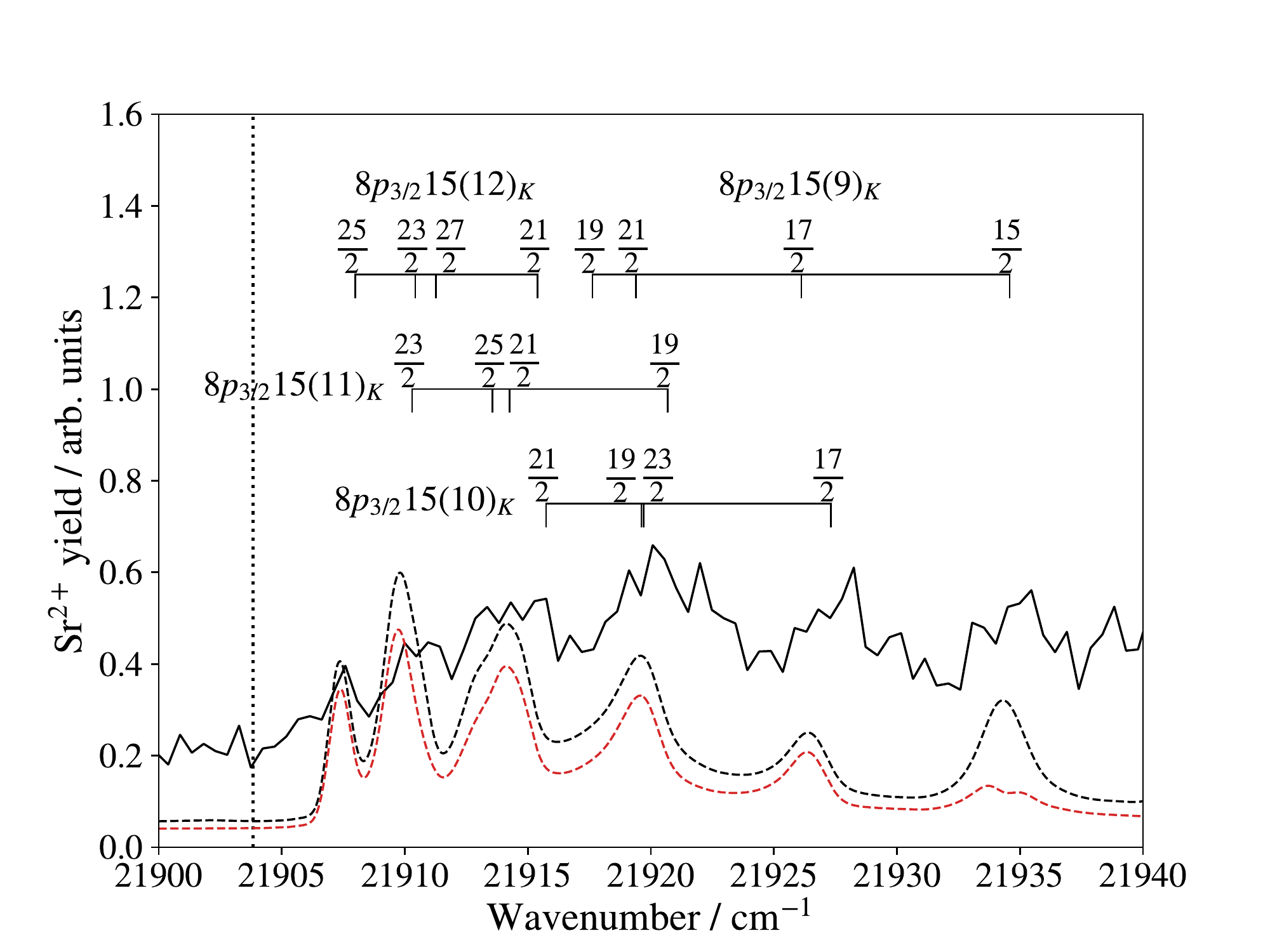}
	\caption{Experimental (full line) and theoretical (dotted lines) Sr$^{2+}$
	spectra in the region of $8p_{3/2}15(l'_2)$ resonances. The red dotted line
	corresponds to the signal calculated from the partial photoionization cross
	section to the Sr$^+(7d)$ states only, whereas the black dotted line was
	obtained from the total cross section. The vertical dotted line indicates the position of a $8p_{3/2}15(l'_2)$ state with zero quantum defect. The theoretical spectra were scaled by a factor of 4 for clarity.}
	\label{fig:8p_n15_zoom}
\end{figure}

The rich substructure of the $n'_2 \sim 15$ band is shown in Fig.~\ref{fig:8p_n15_zoom}. Agreement between theory and experiment is very good and allows us to assign each peak to one or a few $(8p_{3/2}15(9-12)_{j'_2})_{J'}$ states. As for the $7d12l_2'$ region, the spin-orbit splitting of the Rydberg electron is negligible and states were labeled in $jK$ coupling. The energy differences between different $K$ states of a given $l'_2$ manifold are large, and in fact often larger than the energy difference between states belonging to different $l'_2$ manifolds. Thus, $l'_2$ is no longer even an approximately good quantum number. Inspection of the CI coefficients shows that, for example, states with predominant $8p_{3/2}15(l'_2=9)$ character are mixed with states with $l''_2=7, 10, 11$ and 12 and core configurations $7d_{3/2, 5/2}$, $8p_{1/2}$ and $6(5)_{9/2, 11/2}$.

The signal computed from the partial photoionization cross section to
Sr$^+(7d_{3/2, 5/2})$ states \emph{only} is shown by the dotted red line in Fig.~\ref{fig:8p_n15_zoom}, and appears very similar to the signal calculated from the total cross section (black dotted line). Therefore, the doubly-excited $8p_{3/2}15(l'_2)_K$ states autoionize predominantly into $7d\epsilon l$ continua. The kinetic energy $\epsilon$ of the ejected electron is relatively small because the $8p_{3/2}15(l_2)_K$ states are energetically close to the Sr$^+(7d_{3/2, 5/2})$ thresholds ($\sim 200$~cm$^{-1}$). In contrast, autoionization of the $7dnl$ states observed in Fig.~\ref{fig:7d_region} yields higher electron energies because these states are more than 4000~cm$^{-1}$ higher than the closest-lying Sr$^+(7p, 5f)$ thresholds. Autoionization typically becomes less efficient the larger the momentum transfer between the Rydberg and core electrons~\cite{lehec21,genevriez19b}, which translates into the fact that, as observed in the spectra, the overall widths of $7dnl$ states are narrower than the $8pnl$ ones.

Let us now investigate the intensity distribution of the lines in the spectra shown in Fig.~\ref{fig:8p_region}. In contrast with the $7dnl$ region shown in Fig.~\ref{fig:7d_region}, ICE from the $5d_{5/2}$ state to the $8p_{3/2}$ state is allowed by dipole selection rules. We calculated the typical predictions of the independent-electron ICE model~\cite{cooke78a,bhatti81} considering an isolated Rydberg series with a quantum defect of -0.21 and linewidths given by $50 \times 10^{3} / n^3$~cm$^{-1}$, as shown by the dashed gray line in the lower panel of Fig.~\ref{fig:8p_region}. The strongest transition around $21970$~cm$^{-1}$ corresponds to the situation where the principal quantum number of the outer electron is approximately conserved upon photoexcitation ($n_2' \sim n_2 \sim 16$). Shake-up and shake-down of the outer electron are possible and the intensity of the corresponding satellite lines decreases following the $\text{sinc}^2$ law in Eq.~\eqref{eq:overlap_integral}. The ICE model reproduces the overall intensity behavior of the spectrum, however closer inspection reveals that the lines in the theoretical and experimental spectra do not decay as rapidly as the ICE predictions, in particular in the low wavenumber side. It is possible that such a discrepancy is caused by variations of the laser pulse energy with the wavenumber or changes in the detection efficiency with the principal quantum number of the Rydberg electron. However, the low-wavenumber region also corresponds to low principal quantum numbers of the outer electron, where the strongest correlations with the core electron are expected and thus where the largest deviations from ICE should occur. The continuous background in the experimental and theoretical spectra is not reproduced by the ICE model and corresponds to direct ionization into $7d\epsilon l$ continua. It is in fact the continuation of the Rydberg series observed in Fig.~\ref{fig:7d_region} to positive electron energies.  

\subsection{Electronic correlations}\label{sec:electronic_structure}

\begin{figure}
	\includegraphics[width=\columnwidth]{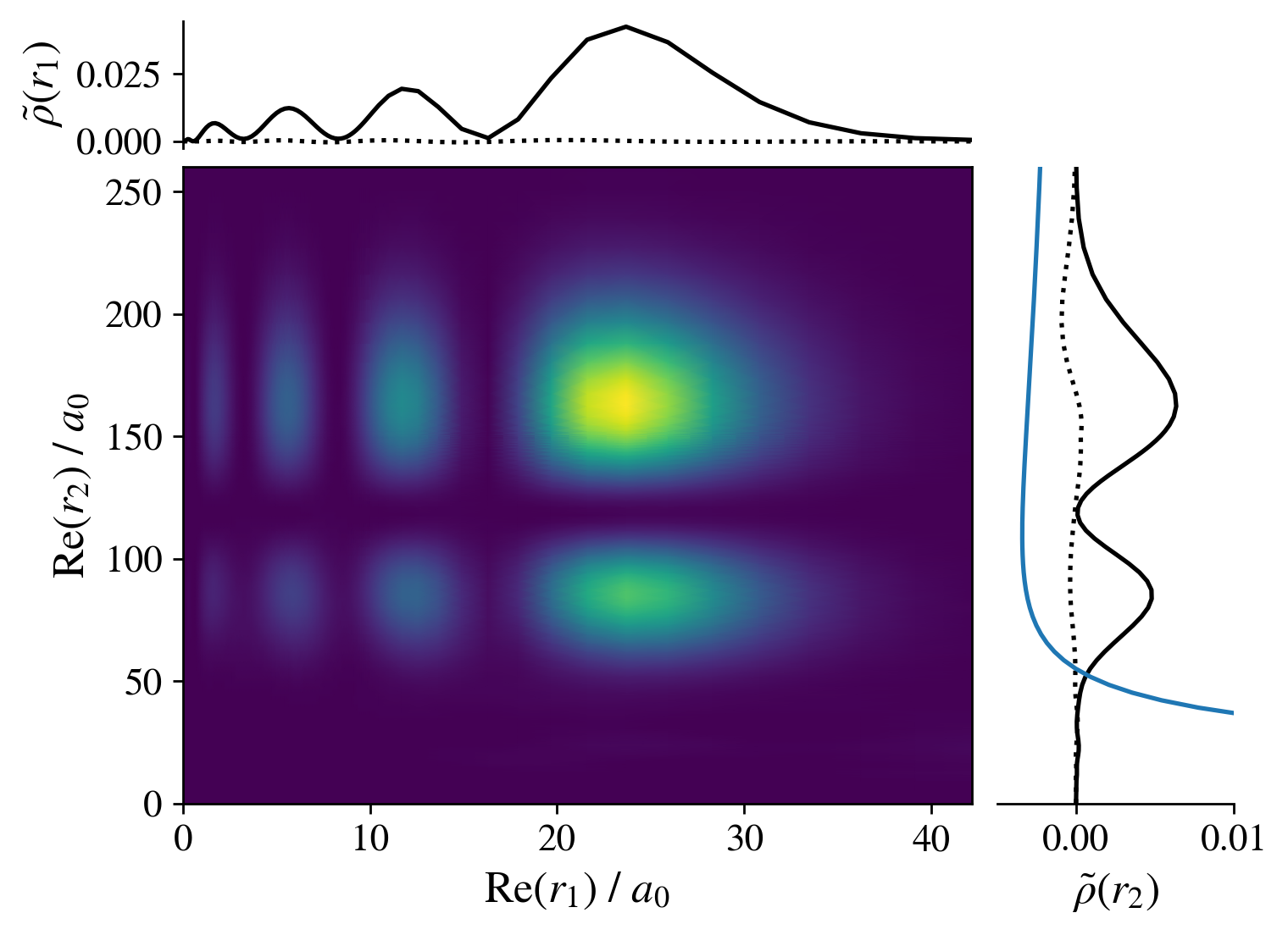}
	\caption{Radial electronic density $\rho(r_1, r_2)$ of the $(7d_{5/2}12(10)_{21/2})_8$ state. Radial densities integrated over $r_2$ and $r_1$ are shown in the top and right panels, respectively. The blue line in the right panel shows the sum of a Coulomb potential and a centrifugal potential for $l = 10$.}
	\label{fig:radial_density}
\end{figure}

\begin{figure}
	\includegraphics[width=\columnwidth]{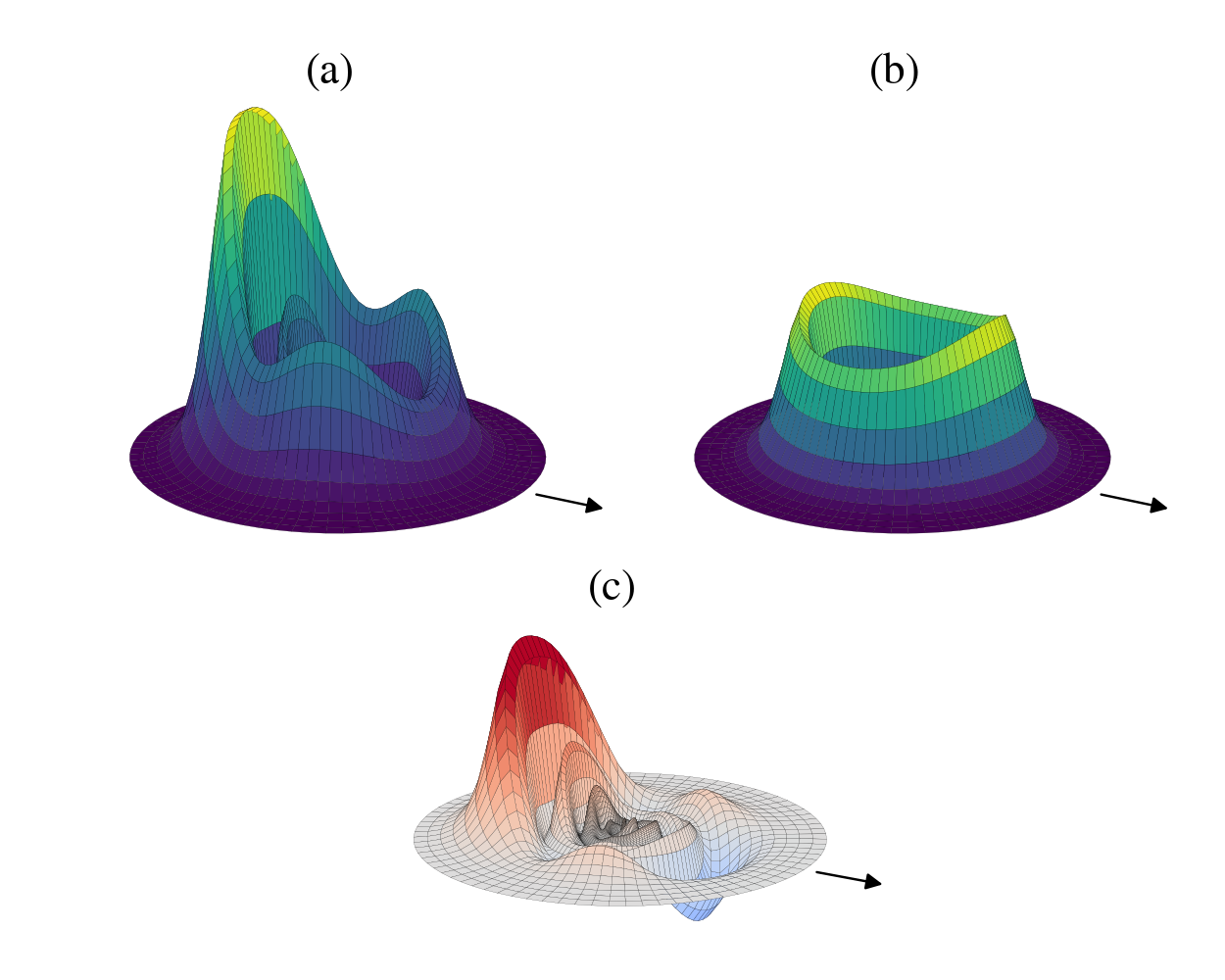}
	\caption{Conditional density $\rho(r_1, \theta_{12} | r_2 = 85 a_0)$ of the $(7d12(10)_{21/2})_8$ state in the presence (a) and absence (b) of electron correlations. Panel (c) shows the difference between the two densities. The black arrow shows the direction along which $\theta_{12} = 0^\circ$.}
	\label{fig:core_density_K15/2}
\end{figure}

\begin{figure}
	\includegraphics[width=\columnwidth]{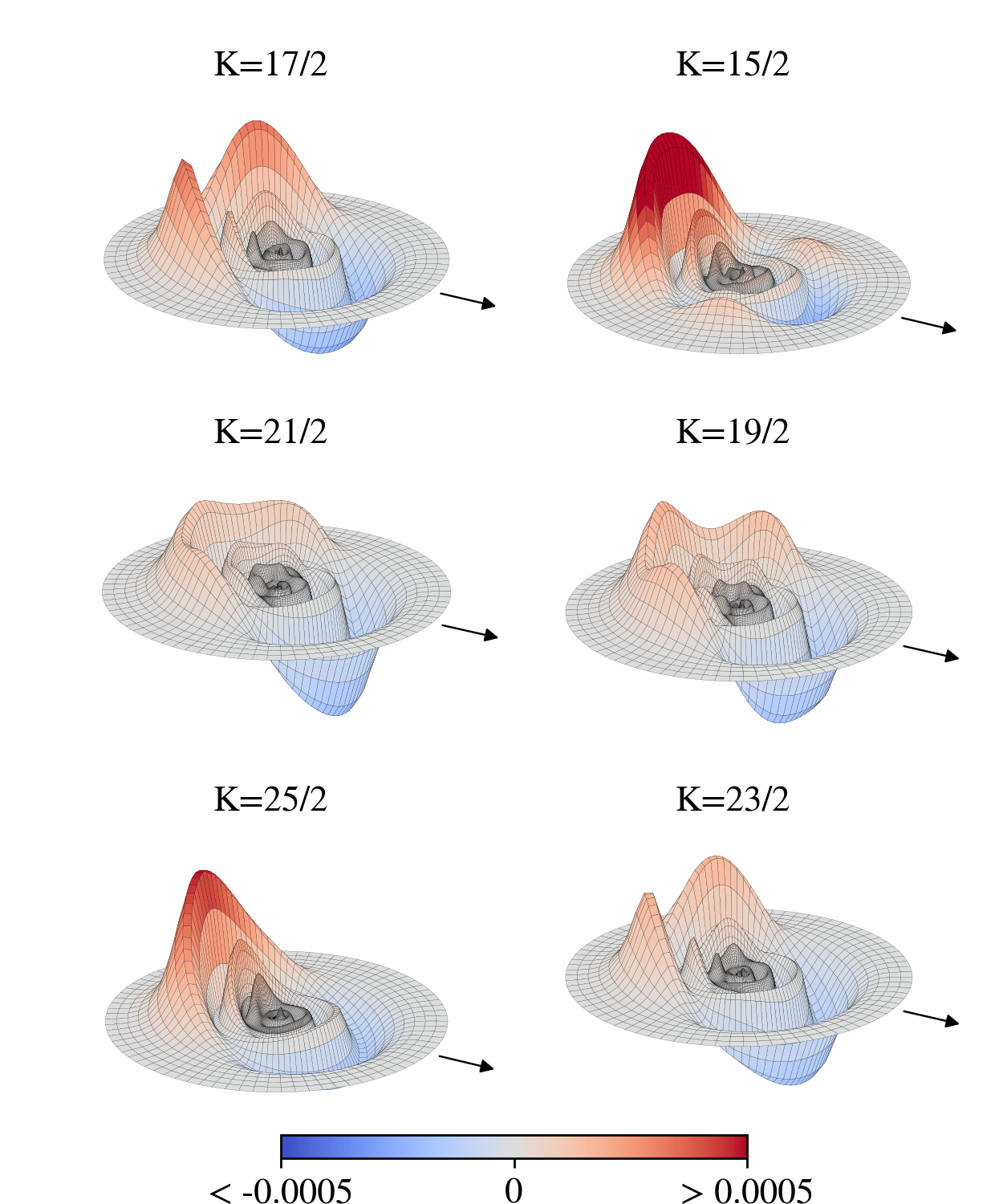}
	\caption{Difference between the conditional probability densities $\rho(r_1, \theta_{12} | r_2 = 85 a_0)$ with and without electronic correlations for the various $(7d_{5/2}12(10))_K$ states. The black arrow shows the direction along which $\theta_{12} = 0^\circ$.}
	\label{fig:core_densitydiff_Ks}
\end{figure}

Electron correlations responsible for the complex structures observed in the
experimental and theoretical spectra above can be investigated in their full
extent with the aid of the two-electron density $\rho(r_1, r_2, \theta_{12})$, calculated from the CI-ECS wavefunctions as described in Sec.~\ref{sec:theory}. We will focus in the following on the $(7d_{5/2}12(10)_{j'_2})_{J'}$ and $(8p_{3/2}13(11)_{j'_2})_{J'}$ states, however we verified that similar conclusions apply to other states as well.

Let us first investigate radial correlations. To do so, we integrate the 3-dimensional two-electron density over the angular coordinate $\cos \theta_{12}$,
\begin{equation}
	\tilde{\rho}(r_1, r_2) = \int \mathrm{d}(\cos\theta_{12})\, \rho(r_1, r_2, \cos\theta_{12}) .
\end{equation}
It yields the radial density shown in Fig.~\ref{fig:radial_density} for the $(7d_{5/2}12(10)_{21/2})_{8}$ state, which is associated with a $K$ value of $15/2$ in $jK$ coupling. Further integration over $r_1$ yields the one-dimensional density whose real (full line) and imaginary (dotted line) parts are shown in the right panel. Integration over $r_2$ gives the density shown in the top panel. The blue line in the right panel shows the sum of a pure Coulomb potential and a centrifugal potential for $l=10$. Complex rotation is applied from $r_{1, 2} = 150\ a_0$ onward, at which point the densities calculated from the CI-ECS wavefunctions do not necessarily follow the usual probabilistic meaning of Hermitian quantum mechanics. Due to exchange symmetry, $\tilde{\rho}(r_1, r_2)$ presents the same behavior for $r_1$ and $r_2$ reversed. This is not shown for clarity.

The radial density in Fig.~\ref{fig:radial_density} reveals that radial
electronic correlations~\cite{fano83} are weak. Indeed, the total density is essentially the product of two independent one-electron densities: one associated with the outer electron density, which resembles the norm squared of the radial function of a $n=12$, $l=10$ Rydberg electron; the other associated with the density of the core electron, which resembles the norm squared of the Sr$^+(7d_{5/2})$ wave function. The absence of radial correlations is not surprising since the two electrons occupy largely different regions of phase space. As shown in the right panel of Fig.~\ref{fig:radial_density}, the radial
density associated with the outer electron is confined to $r \gtrsim
50~a_0$ by the large centrifugal barrier it experiences, whereas the inner electron density vanishes at $r \sim 40\ a_0$ and beyond. 


Angular correlations~\cite{fano83}, on the other hand, are far from negligible.
Figure~\ref{fig:core_density_K15/2}(a) shows the conditional density of the core electron when the outer electron is fixed at $r_2 = 85\ a_0$, corresponding to the first maximum of the outer-electron distribution in
the right panel of Fig.~\ref{fig:radial_density}. One notices that the inner electron preferentially sits at $\theta_{12}=180^\circ$, \textit{i.e.}, on the side of the nucleus opposite to the outer electron. Comparison with the conditional density calculated using the independent-electron approximation, shown in Fig.~\ref{fig:core_density_K15/2}(b), reveals that the strong polarization of the core-electron density is caused by its interaction with the outer electron. In order to make correlation-induced changes in the core-electron density more apparent, we also calculated the difference between the conditional densities with (Fig.~\ref{fig:core_density_K15/2}(a)) and without (Fig.~\ref{fig:core_density_K15/2}(b)) correlations, as shown in Fig.~\ref{fig:core_density_K15/2}(c).

The differences between the conditional densities $\rho(r_1, \theta_{12} | r_2 = 85\ a_0)$ with and without correlations are shown in Fig.~\ref{fig:core_densitydiff_Ks} for the various $7d_{5/2}12(10)_K$ states. In all cases, the inner-electron density is significantly distorted by its interaction with the outer electron. The densities for the different $K$ values can be grouped in three categories with $K=15/2$ and $25/2$, $K=17/2$ and $23/2$, and $K=19/2$ and $21/2$, respectively. They are characterized by the presence of 0, 1 and 2 minima in the conditional density along the $\theta_{12}$ direction in the $[\frac{\pi}{2}, -\frac{\pi}{2}]$ half-plane, respectively. In the limit where the radial motion of the outer electron can be adiabatically decoupled from that of the inner electron, these three categories correspond to different projections of the total orbital angular momentum of the core electron onto an adiabatic quantization axis defined by the distance between the nucleus and the outer electron ($\bm{r}_2$).

\begin{figure}
	\includegraphics[width=\columnwidth]{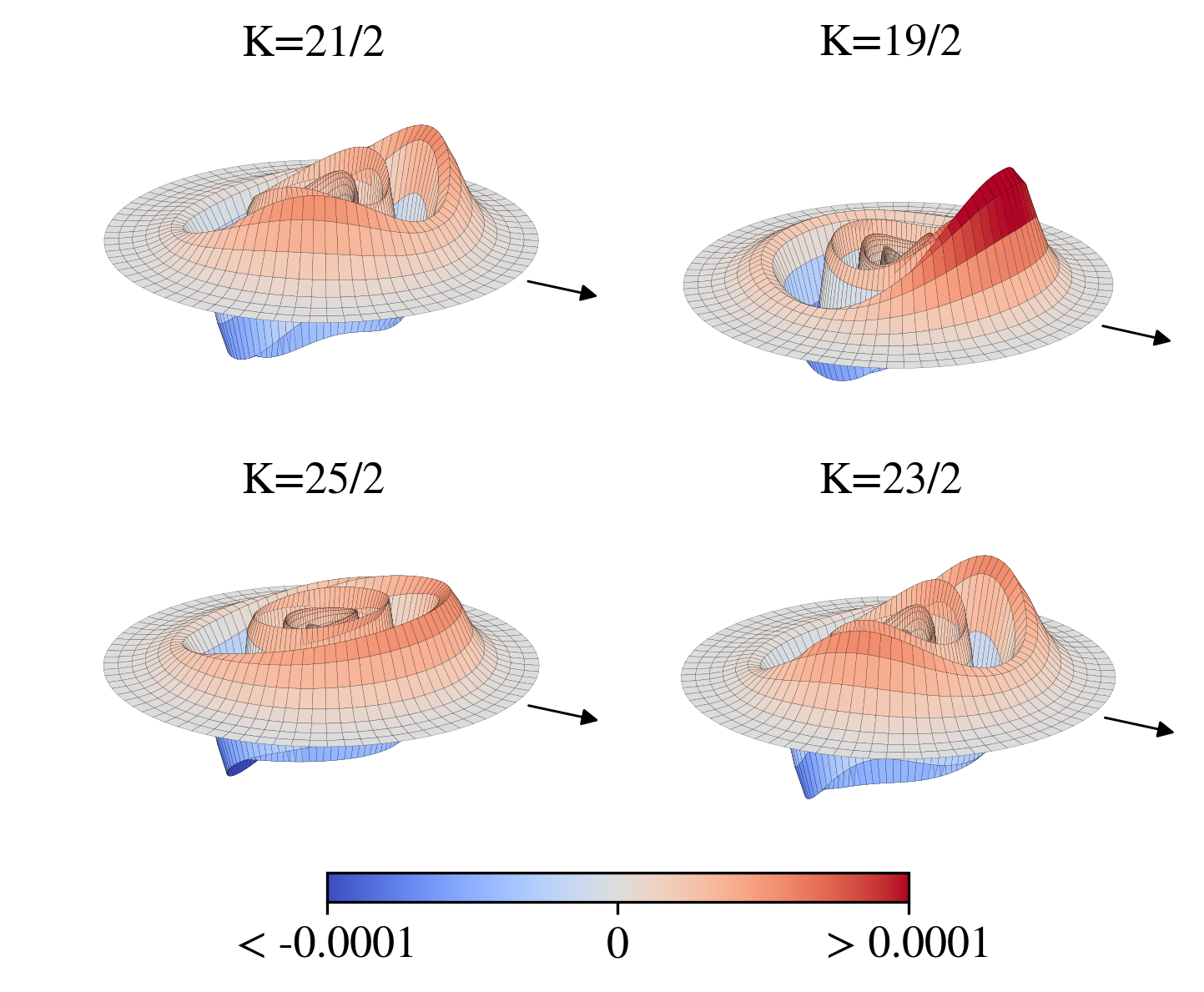}
	\caption{Difference between the conditional probability densities $\rho(r_1, \theta_{12} | r_2 = 120\ a_0)$ with and without electronic correlations for the various $(8p_{3/2}13(11))_K$ states. The black arrow shows the direction along which $\theta_{12} = 0^\circ$.}
	\label{fig:core_densitydiff_Ks_8p}
\end{figure}

Figure~\ref{fig:core_densitydiff_Ks_8p} shows the differences between the
conditional densities $\rho(r_1, \theta_{12} | r_2 = 120\ a_0)$ with and
without correlations for the various $8p_{3/2}13(11)_K$ states. Contrary to the $7d_{5/2}12(10)_K$ states, the core-electron density is polarized toward $\theta_{12} = 0^\circ$, as is particularly visible for $K=19/2$. The inner electron thus preferentially resides on the same side of the nucleus as the outer electron.
The densities can be grouped into $K=19/2$ and $25/2$, and $K=21/2$ and
$23/2$, characterized by the presence of 0 and 1 minima along the $\theta_{12}$ direction in the $[-\frac{\pi}{2}, \frac{\pi}{2}]$ half-plane, respectively.

\section{Discussion}\label{sec:discussion}

Because the effective principal quantum number of the outer electron is
significantly larger than that of the inner one, the classical orbit period of
the former is much larger than the one of the latter ($n_2^3 \gg n_1^3/4$). The
precession of the Runge-Lenz vector, which is classically equivalent to the
precession of major axis associated to the elliptic orbit of the inner electron,
is also much faster than the classical orbit period of the outer electron
because~\cite{eichmann92} \begin{equation} n_2^3 \gg
\frac{n_1^3}{4}\left(\frac{\partial\mu_{l_1}(E)}{\partial l_1}\right)^{-1} ,
\label{eq:adiabaticity_condition} \end{equation}  
represents the quantum defects of Sr$^+(n_1l_1)$ orbitals at energy $E$.  Thus,
in first approximation, the radial motion of the outer electron can be
adiabatically decoupled from the motion of the inner electron (see,
\textit{e.g.}, Refs.~\citenum{heber97,ostrovsky95,richter92}). Such a procedure
is further supported by the absence of strong radial correlations in the CI-ECS
two-electron densities. Within the adiabatic picture, the inner electron evolves
in the static electric field generated by the outer electron, which is fixed at
a given point $\bm{r}_2$ in space. This approach is equivalent to the
clamped-nuclei approximation in molecular physics, and offers a qualitative
understanding of the quantum defects and charge densities presented above. In
the Sr$^+$ ion, the scalar polarizability of the $7d_{5/2}$ orbital, calculated
using the one-electron radial functions of Sec.~\ref{sec:CI-ECS}, is $\alpha_0
\sim 54332\ a_0^3$. Thus, the Sr$^+(7d_{5/2})$ state is red-shifted by its
interaction with an external electric field and the charge distribution
localizes preferentially on the side of the nucleus opposite to the electric
field direction. In the present case, this translates into the core electron
sitting preferentially on the side of the nucleus opposite to the outer
electron. Conversely, the scalar polarizability of Sr$^+(8p_{3/2})$ is
$\alpha_0\sim -58794\ a_0^3$. It is blue shifted by an external electric field
and the center of gravity of the charge distribution is displaced away from the
nucleus in the direction of the electric field. The inner electron
preferentially resides on the same side of the nucleus as the outer electron.
The fact that the polarizabilities of Sr$^+(7d_{5/2})$ and Sr$^+(8p_{3/2})$ have
similar magnitudes but different signs further explains why $7p_{5/2}nl$ and
$8p_{3/2}nl$ doubly-excited states exhibit quantum defects of similar magnitude
but opposite signs. For $7d_{5/2}(n'_2=12, l'_2=10)$ states, the electric field generated by the
outer electron at its mean radial distance induces an energy shift due to the
scalar polarizability of the ion core of $~\sim -31$~cm$^{-1}$, comparable to the shift observed
in the experimental spectrum. For $8p_{3/2}(n_2'=15, l_2'=11)$ states, the shift is $\sim
12$~cm$^{-1}$, which compares well with observations (see
Fig.~\ref{fig:8p_n15_zoom}). Energy shifts depending on the projection of the
total angular momentum of the inner electron onto the adiabatic quantization axis
($\bm{r}_2$) can in principle be derived by taking into account the tensor
polarizability ($\alpha_2$) of the Sr$^{+}$ orbital under
consideration~\cite{mitroy10}. However, the agreement with the observed and
calculated energies is not significantly improved, a fact we attribute primarily
to nonadiabatic effects not accounted for in our simple model.

The doubly-excited states under scrutiny are called planetary
states~\cite{percival77,eichmann90,camus94,eichmann92}, by analogy with planets in a solar system. The
autoionization lifetimes we calculate for these states are long ($\sim 10^{-12}$~s$ - 10^{-10}$~s for $7d_{5/2}12(10)$ states) compared to the classical orbit period of the Rydberg
electron ($\tau \sim 40$~fs for $n=12$). The planetary system is thus metastable and survives for a large
number of orbits. The two electrons describe a synchronous angular motion whose
details depend primarily on the polarizability of the ion core and the projection of the
ion-core total angular momentum onto the adiabatic outer-electron-nucleus axis.
For the $7pnl$ and $8pnl$ double-Rydberg states considered in the present article, the two electrons are preferentially
either at $0^\circ$ or $180^\circ$ from each others depending on the sign of
the static polarizability of the ion core. However, their radial motion is not
constrained and they are free to roam along the electron-nucleus-electron
direction. This differs from the frozen planet states calculated by Richter and
Wintgen for low-angular-momentum, high-lying doubly-excited states of helium~\cite{richter91,richter92}, in which the two electrons are further localized at specific $r_1$ and $r_2$ values on the radial axis due to strong radial correlations.


\section{Conclusion}

We presented an experimental and theoretical study of $7dnl$ and $8pnl$ planetary states of Sr. Experimental spectra were recorded by photoexcitation from $5dnl$ states, prepared by multiphoton excitation of ground-state atoms and using the Stark-switching technique, followed by double-ionization \textit{via} autoionization, photoionization, field-ionization and any combination thereof. Theoretical spectra were calculated using the method of configuration interaction with exterior complex scaling, which permitted the complete treatment of the dynamics of the two highly-excited electrons from first principles. Good agreement was obtained with experimental data, and permitted the detailed analysis of the complex structures observed in the spectra. The signatures of electron-electron correlations in the spectra were carefully investigated. Electronic correlations were further investigated using two-electron conditional probability densities calculated from CI-ECS wavefunctions. A strong distortion of the inner-electron density caused by its interaction with the outer electron was observed, and was related to the polarization of the ion core by the electric field of the outer electron and to the orientation of the ion-core angular momentum relative to the nucleus-outer-electron adiabatic axis. The present work thus validates, through calculations from first principles in quantitative agreement with high-resolution experimental results, the frozen-planet approximation that describe qualitatively the structure of planetary states based on the polarization of the inner electron by the Rydberg electron~\cite{camus89,jones90,eichmann90,eichmann92,heber97}. Similar ideas were developed in the early days of quantum mechanics~\cite{mayer33,born24} and are being used, \textit{e.g.}, to study high-$l$ singly-excited Rydberg series of K~\cite{peper19}. In doubly-excited Rydberg states, core-polarization effects are dramatically enhanced compared to singly-excited Rydberg states because static polarizabilities scale as $n^7$~\cite{gallagher94}. We further see that quantitative agreement between calculated and measured energy levels can only be reached through complete calculations taking into account higher order electrostatic interactions and nonadiabatic effects.

For larger orbital momenta $l_1$ of the inner electron, the ion core is expected to be more polarizable and the dynamics described in the present work should be yet more pronounced. Such states are for example the $6gnl$ or $6hnl$ states lying energetically just above the $8pnl$ states we studied. Moreover, because the quantum defects $\mu_{l_1}$ associated with high $l_1$ values are small, Eq.~\eqref{eq:adiabaticity_condition} should be violated for values of the principal quantum number $n_2$ of the outer electron that are accessible to both theory and experiment. The study of such states is a perspective of future work as it would provide further information on the dynamics of planetary states. It would also allow one to track the progressive breakdown of the frozen-planet approximation as, for decreasing $n_2$ values, the motion of the outer electron can no longer be adiabatically separated from the motion of the inner electron and radial correlations become more pronounced. Another perspective is the study of states where the centrifugal barrier associated with the outer electron is small enough to allow its penetration in the core region. Frozen planet states~\cite{richter91}, exhibiting strong radial \emph{and} angular correlations, are known to form under such conditions in the He atom but their existence in other systems remains elusive. 

\begin{acknowledgments}
	The authors gratefully acknowledge fruitful discussions with F.\,Merkt and D.\,Wehrli and W.\,Huang for his help with the experiment at an early stage of this work.
\end{acknowledgments}

\bibliographystyle{apsrev4-1}
\bibliography{paper}

\end{document}